\documentclass[11pt,english]{article}
\usepackage{amsmath}
\usepackage{amsthm}
\usepackage{newtxtext,newtxmath}
\usepackage[T1]{fontenc}
\usepackage[latin9]{inputenc}
\usepackage{geometry}
\usepackage{babel}
\usepackage{float}
\usepackage{url}
\usepackage{graphicx}
\usepackage[labelfont=bf,labelsep=period]{caption}
\usepackage{subcaption} % Only this one for subfigures

\usepackage{booktabs}
\usepackage{threeparttable}

\usepackage{setspace}
\usepackage[authoryear]{natbib}
\setcitestyle{round}
\makeatletter
\def\NAT@forcefull#1{\@ifstar{#1}{#1}}
\def\citet{\begingroup\NAT@swafalse\let\NAT@ctype\z@\NAT@partrue
  \NAT@forcefull{\NAT@fulltrue\NAT@citetp}}
\def\citep{\begingroup\NAT@swatrue\let\NAT@ctype\z@\NAT@partrue
  \NAT@forcefull{\NAT@fulltrue\NAT@citetp}}
\def\cite{\begingroup\let\NAT@ctype\z@\NAT@partrue\NAT@swatrue
  \NAT@forcefull{\NAT@fulltrue\NAT@cites}}
\def\citealt{\begingroup\NAT@swafalse\let\NAT@ctype\z@\NAT@parfalse
  \NAT@forcefull{\NAT@fulltrue\NAT@citetp}}
\def\citealp{\begingroup\NAT@swatrue\let\NAT@ctype\z@\NAT@parfalse
  \NAT@forcefull{\NAT@fulltrue\NAT@citetp}}
\makeatother
\usepackage[unicode=true,pdfusetitle,
 bookmarks=true,bookmarksnumbered=false,bookmarksopen=false,
 breaklinks=false,pdfborder={0 0 0},pdfborderstyle={},backref=false,colorlinks=true]
 {hyperref}
\hypersetup{
 linkcolor=red,urlcolor=red,citecolor=blue}

\usepackage{eurosym}

\makeatletter

\usepackage[bottom]{footmisc}
\usepackage{indentfirst}
\usepackage{lscape}
\usepackage{xcolor}
\usepackage{titlesec}

\date{July 2026}

\makeatother

\begin{document}
\title{Does Regulation Bite at Gateways? Evidence from MiCA and Stablecoins\footnote{Nicola Borri is with LUISS University, Rome. Kirill Shakhnov is with University of Surrey.}}
\author{Nicola Borri, and Kirill Shakhnov}

\maketitle
\begin{abstract}
\noindent Gateways are trading venues where regulation can change the assets investors can trade. We study this margin using MiCA--- EU's Markets in Crypto-Assets Regulation---which led several exchanges to delist USDT pairs for European Economic Area users, while USDC obtained MiCA authorization. First, aggregate market shares and trading volumes barely move. Second, comparing Regulated-facing exchanges with globally oriented exchanges where MiCA is less likely to bind, we show that the cross-section shifts toward USDC---USDC share rises by 0.82 and relative trading volume by 0.54 pre-event standard deviations. Both reflect USDT trading contracting where it is delisted.

\end{abstract}
\thispagestyle{empty}

\newpage
%\pagebreak{}

\setcounter{page}{1}

\section{Introduction}

Gateways are trading venues and other intermediaries where regulation can change the assets investors can trade, even when activity can migrate beyond a jurisdiction's reach. We study this margin using the European Union's Markets in Crypto-Assets Regulation (MiCA). The European Union is a large economic area---about 17 percent of world GDP and approximately 10 percent of world equity-market capitalization \citep{SIFMA2025Factbook}---and its crypto market is sizable: European users received more than \$2.6 trillion in crypto value in the year to mid-2025, second only to North America \citep{Chainalysis2025Index}. MiCA created a sharp asymmetry between the two dominant stablecoins: USDT, the largest by market capitalization, was not authorized; USDC, the second largest, was. Several exchanges therefore removed USDT pairs for European Economic Area (EEA) users; the exchanges that did so account for 50 percent of pre-event USDT--USDC trading volume in our sample.\footnote{Using aggregate exchange volume across all pairs---not limited to stablecoins---Regulated-facing exchanges account for approximately 9 percent of sample volume.} We take April 1, 2025 as the common event date and study two exchange-level outcomes, the USDC share of USDT--USDC trading and the volume of USDC trading relative to USDT, both of which measure substitution toward USDC. We compare exchanges facing EU users---\textit{Regulated-facing exchanges}---with globally-oriented exchanges where MiCA is less likely to bind---\textit{Global exchanges}.

Stablecoin regulation is an ideal laboratory for this question. It changes the set of assets traded at regulated gateways while leaving users able to trade around the clock, hold accounts at multiple exchanges, and move crypto-assets across venues at low cost \citep{MakarovSchoar2020,borri2022crosssection}. We start from CoinMarketCap's top-30 centralized-exchange ranking and retain the exchanges for which the USDT--USDC panel can be constructed, a 14-exchange sample that accounts for 57 percent of aggregate top-30 trading volume. Using audience data from Similarweb, a web-traffic analytics provider, we label an exchange \textit{Regulated-facing} when its EU audience share exceeds 10 percent, and \textit{Global} otherwise: four exchanges are Regulated-facing and ten are Global.

The empirical analysis has three main findings. First, aggregate markets absorb the regulatory shock. Over the 15 days before and after April 1, USDC's share of combined USDT--USDC trading volume moves only from 17.70 to 18.24 percent. Aggregate trading also moves little in relative terms: USDC volume rises from 1.18 to 1.91 billion dollars per day and USDT volume from 5.46 to 8.56 billion, leaving the aggregate USDC/USDT volume ratio near flat, from 0.215 to 0.223. This contrasts with earlier crypto-regulation episodes, where shocks such as China's 2017 exchange ban produced large aggregate effects \citep{borri2020regulation}.

Second, stablecoin trading shifts toward USDC on Regulated-facing exchanges. The outcome is each exchange's USDC share of USDT--USDC trading, standardized and detrended using only pre-event data. After April 1, the USDC share rises on Regulated-facing exchanges relative to Global exchanges by a difference-in-differences of 0.82 pre-event standard deviations (exchange-clustered $p{=}0.004$), about 6 percentage points in raw terms. The two groups move in opposite directions: the standardized share falls by 0.24 on Global exchanges and rises by 0.58 on Regulated-facing exchanges, from pre-event levels of roughly 7 percent on Global venues and 46 percent on Regulated-facing exchanges.

Third, the volume evidence shows that this shift works by a contraction in USDT rather than by expanding USDC on Regulated-facing exchanges. A rising USDC share can reflect either more USDC trading or less USDT trading. Measured as the log ratio of USDC to USDT volume, standardized and detrended within exchange, the difference-in-differences is 0.54 pre-event standard deviations (exchange-clustered $p{=}0.043$), so relative volume moves toward USDC on Regulated-facing exchanges. The decomposition attributes this to USDT: its volume falls by about 20 percent on Regulated-facing exchanges relative to Global exchanges (exchange-clustered $p{=}0.003$), while USDC volume does not change significantly (about 4 percent, $p{=}0.73$). At the same time, USDT volume grows on Global venues, so aggregate ratios can remain nearly flat.

Finally, we also show preliminary evidence that the reallocation affects market prices. A stablecoin is meant to trade at par with the dollar---one token for one dollar, its \emph{peg}---but its price can slip below par under selling pressure. After April 1, the USDT price slips below par by slightly more than USDC on Regulated-facing exchanges, and the sign is consistent across all four Regulated-facing exchanges. The effect is small---a few basis points---and estimated imprecisely, because deviations from parity are tiny and the sample is small.

In the Online Appendix, we show that the share and volume results are robust to a range of implementation choices. We present alternative smoothing windows, a raw percentage-point USDC-share specification, leave-one-out diagnostics, wild cluster bootstrap inference, exact randomization inference, and event-date robustness based on exchange-specific USDT delisting dates.

This paper contributes to two strands of the literature. The first studies how regulation bites when activity can evade, relocate, or arbitrage across regimes. Prior work studies evasion and enforcement in taxation \citep{AllinghamSandmo1972,KlevenKnudsenKreinerPedersenSaez2011,Pomeranz2015}, tax-haven leakage \citep{Zucman2013}, regulatory arbitrage and cross-border spillovers in banking and securities markets \citep{Stigler1971,HoustonLinMa2012,OngenaPopovUdell2013,Plantin2015}, and bonding through stricter intermediaries \citep{DoidgeKarolyiStulz2004}. We bring this question to crypto markets, where migration across venues is easiest and intermediaries differ sharply in regulatory reach.
The second strand studies cryptocurrency regulation and digital-asset markets. Existing work examines policy shocks, market frictions, liquidity provision, trading volumes, spillovers across crypto markets, and segmentation across jurisdictions \citep{borri2020regulation,borri2023cryptomarket,borri2025coming,copestake2023crypto,dufouleur2026bitcoin,farag2025returns}. We show that gateway regulation redirects stablecoin trading across exchanges while leaving aggregates essentially unchanged.

\section{Institutional Setting, Data, and Aggregate Evidence}
\label{sec:mica_data}

MiCA (Regulation (EU) 2023/1114) creates a single EU rulebook for crypto-assets not already covered by existing financial services legislation \citep{EuropeanParliamentCouncil2023MiCA,ESMA2025MiCA}. It distinguishes asset-referenced tokens (ARTs), e-money tokens (EMTs)---the category that covers dollar-pegged stablecoins such as USDT and USDC---and other crypto-assets, with stronger reserve, redemption, and supervisory requirements for stablecoin-like instruments. Crypto-asset service providers (CASPs) must be authorized and can then passport services across the EU. MiCA entered into force on June 29, 2023; the ART/EMT regime applied from June 30, 2024; the broader regime for other crypto-assets and CASPs applied from December 30, 2024. 

The provision that drives our design is issuer authorization: Tether did not seek it for USDT, while Circle's USDC was authorized. Appendix~\ref{app:mica_details} provides additional institutional detail. Exchanges serving EEA users therefore removed USDT trading pairs (for details, see Appendix Table~\ref{tab:app_delistings}). We take April 1, 2025 as the common event date because it immediately follows the March 31 delisting wave in our sample.

Our exchange universe starts from the CoinMarketCap top-30 centralized exchanges. We retain venues where both USDT and USDC trade and where pair-level data support a daily panel; this screening yields 14 exchanges observed daily from January 1, 2024 to December 7, 2025. Pair-level trading volume comes from CryptoCompare.

We classify each exchange by its Similarweb web-audience shares. An exchange is \textit{Regulated-facing} if its EU audience share exceeds 10 percent, and \textit{Global} otherwise. The four Regulated-facing exchanges are Bitstamp, Coinbase, Gemini, and Kraken; each also draws more than 10 percent of its audience from the US, so all four face both EU and US users. The ten Global exchanges are Binance, BitMart, Bitfinex, Bitget, Bybit, Gate, HTX, KuCoin, MEXC, and OKX. All four Regulated-facing exchanges removed USDT pairs for European users during our window. The classification targets visibility rather than legal exposure. Binance also removed USDT pairs for EEA users on March 31, 2025, but remains in Global because non-US and non-EU traffic dominates its volume.

\begin{figure}[H]
    \centering
    \includegraphics[width=0.92\linewidth]{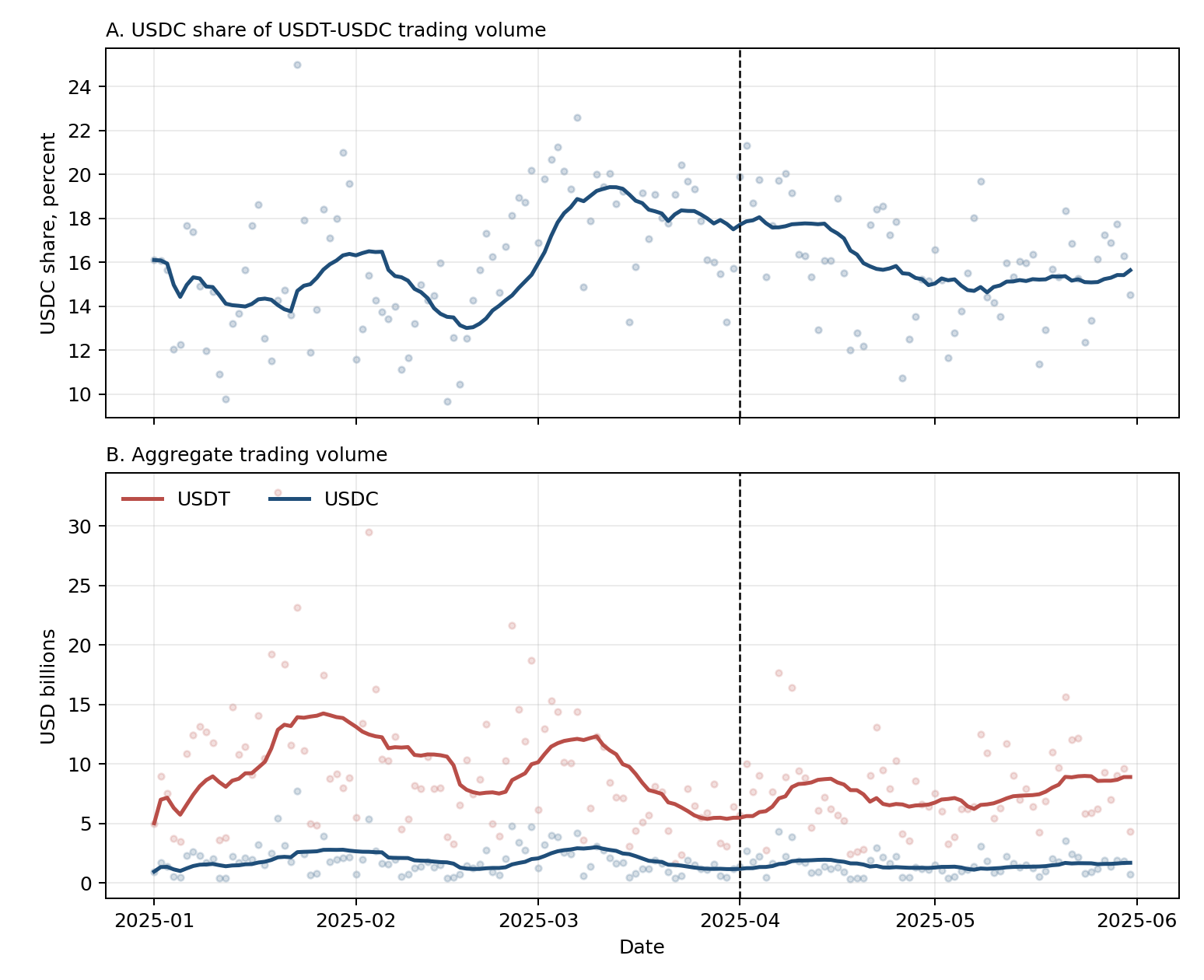}
    \caption{Aggregate Stablecoin Trading Share and Trading Volume}
    \label{fig:aggregate_markets}
    \begin{center}
    \parbox{0.95\linewidth}{\scriptsize \textit{Note:} Panel A plots daily aggregate USDC trading share, defined as USDC trading volume divided by USDC plus USDT trading volume, summed across the 14 exchanges in the sample. Panel B plots daily USDT and USDC trading volume summed across the same exchanges. Solid lines are 14-day trailing averages; dots are daily observations. The vertical dashed line corresponds to April 1, 2025, the cutoff date we use to denote the introduction of the MiCA regulation. Volumes are in USD billions.}
    \end{center}
\end{figure}

Figure~\ref{fig:aggregate_markets} shows the aggregate trading outcomes. Panel A plots USDC's share of combined USDT--USDC trading volume, the aggregate counterpart of the first exchange-level outcome; Panel B plots USDT and USDC trading volume summed across the 14 exchanges. Over the 15 days before and after the event day, aggregate USDC trading share moves only from 17.70 to 18.24 percent. Trading rises for both coins---USDC from 1.18 to 1.91 billion dollars per day, USDT from 5.46 to 8.56 billion---leaving the aggregate USDC/USDT volume ratio nearly flat, from 0.215 to 0.223. The aggregate is dominated by large Global venues that kept offering USDT, so a reallocation concentrated on Regulated-facing venues is largely invisible in the total. This motivates the exchange-level analysis: if MiCA bites through venue compliance and user access, the effect should appear in the cross-section, not in the aggregate.

\section{Empirical Framework}
\label{sec:framework}

We estimate a difference-in-differences event study around $T^{*}={}$April 1, 2025, over a 30-day event window covering the 15 days before and after the cutoff. Outcomes are constructed in three steps. For each exchange, we take a 14-day trailing moving average of the daily series, standardize it by the exchange's pre-event mean and standard deviation, and remove an exchange-specific linear trend estimated on pre-event observations only. Smoothing filters daily noise, standardization puts exchanges with very different levels and volatilities on a common scale, and detrending removes exchange-specific drift. The specification is
\begin{equation}
y_{it} = \alpha_i + \gamma\, D^{post}_{t} + \beta \left(\mathrm{Reg}_i \times D^{post}_{t}\right) + \varepsilon_{it},
\label{eq:did}
\end{equation}
where $y_{it}$ is the outcome, $\alpha_i$ are exchange fixed effects, $D^{post}_{t}=\mathbf{1}\{t \geq T^{*}\}$, and $\mathrm{Reg}_i$ indicates a Regulated-facing exchange. The coefficient of interest is $\beta$, the post-event change on Regulated-facing exchanges relative to Global exchanges. Observations receive equal exchange-day weights.

Identification rests on parallel trends: absent the delistings, Regulated-facing and Global exchanges would have evolved similarly over the event window. Consistent with this assumption, the raw series---smoothed only by a trailing average, with no standardization or detrending---move together before the event and diverge around the cutoff (see Appendix~\ref{app:pretrends}). Because inference with 14 exchanges is a first-order concern, we report non-clustered, exchange-clustered, and two-way (exchange $\times$ date) clustered standard errors, together with exact randomization inference over all $\binom{14}{4}=1001$ treatment assignments and a wild cluster bootstrap \citep{BertrandDufloMullainathan2004,CameronGelbachMiller2008,MacKinnonWebb2017}.

\section{Results}
\label{sec:results}

We study two outcomes. The first is the USDC share of USDT--USDC trading,
\[
\text{USDC Share}_{it} =
\frac{\text{USDC volume}_{it}}{\text{USDC volume}_{it}+\text{USDT volume}_{it}},
\]
and the second is the relative-volume spread,
\[
\text{Relative Volume}_{it} =
\log(1+\text{USDC volume}_{it})-\log(1+\text{USDT volume}_{it}).
\]
Both capture substitution toward USDC, from different angles: the share measures the composition of an exchange's USDT--USDC trading, while the spread compares trading volumes in the two coins directly and is not bounded between zero and one. Under both, a higher value means USDC gains relative to USDT. Although the two are mechanically related---both rise with the USDC-to-USDT ratio---reporting them together is informative: agreement between a bounded composition measure and an unbounded volume measure guards against either result being an artifact of its functional form, and, because the spread is in trading-volume units, it is the measure we can decompose to see whether the tilt toward USDC comes from more USDC trading or less USDT trading (Appendix Table~\ref{tab:app_volume_legs}).

\begin{figure}[htp]
    \centering
    \includegraphics[width=0.9\linewidth]{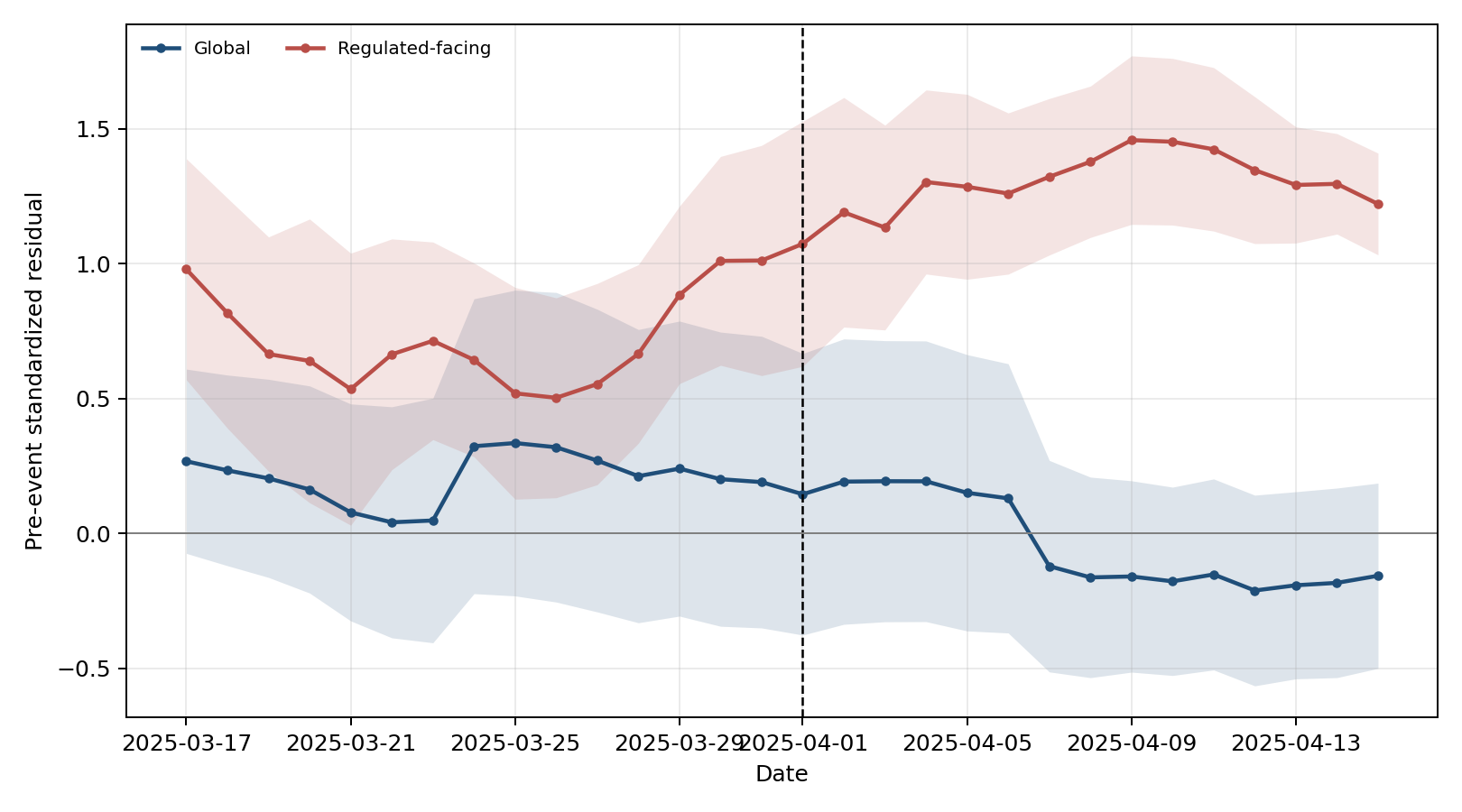}
    \caption{USDC share around MiCA}
    \label{fig:share_path_event}
    \begin{center}
    \parbox{0.95\linewidth}{\scriptsize \textit{Note:} Daily equal-weighted mean USDC-share outcome on Global and Regulated-facing exchanges. For each exchange, the 14-day trailing average of the USDC share is standardized using that exchange's pre-April-1 mean and standard deviation and detrended by an exchange-specific linear time trend estimated on pre-event observations. Global exchanges are Binance, BitMart, Bitfinex, Bitget, Bybit, Gate, HTX, KuCoin, MEXC, and OKX; Regulated-facing exchanges are Bitstamp, Coinbase, Gemini, and Kraken. The vertical dashed line marks April 1, 2025. Shaded bands are $\pm 1$ cross-exchange standard error of the daily group mean.}
    \end{center}
\end{figure}

Figure~\ref{fig:share_path_event} plots the event-window path of the USDC-share outcome. The path is flat to slightly declining on Global exchanges; on Regulated-facing exchanges it rises around April 1, and the separation persists through the end of the window.

Table~\ref{tab:main_event} reports the estimates. In Panel A (USDC share), the outcome falls by 0.24 pre-event standard deviations on Global exchanges and rises by 0.58 on Regulated-facing exchanges, a difference-in-differences of 0.82 (exchange-clustered $p{=}0.004$, two-way $p{=}0.004$). In raw terms this is a relative gain of about 6 percentage points, from pre-event levels of roughly 46 percent on Regulated-facing exchanges and 7 percent on Global venues. Panel B (relative volume) is directionally identical: the DiD is 0.54 standard deviations (exchange-clustered $p{=}0.043$). Decomposing volume into its two legs shows the shift comes from USDT: its volume falls by about 20 percent on Regulated-facing exchanges relative to Global (exchange-clustered $p{=}0.003$), while USDC volume does not change significantly (Appendix Table~\ref{tab:app_volume_legs}). The reallocation reflects USDT contracting where it is delisted, not USDC expanding.

\begin{table}[H]
\centering
\begin{threeparttable}
\caption{Event Study: USDC Share (Panel A) and Relative Volume (Panel B)}
\label{tab:main_event}
\small
\setlength{\tabcolsep}{4pt}
\begin{tabular}{lrrrrrrr}
\toprule
Effect & Estimate & SE non-cl. & $p$ non-cl. & SE exch. & $p$ exch. & SE 2-way & $p$ 2-way \\
\midrule
\multicolumn{8}{l}{\textit{Panel A: USDC share}}\\
Global post-pre change & -0.243 & 0.058 & $<0.001$ & 0.110 & 0.028 & 0.104 & 0.019 \\
Regulated-facing post-pre change & 0.576 & 0.092 & $<0.001$ & 0.261 & 0.028 & 0.258 & 0.026 \\
Regulated-facing $-$ Global DiD & 0.818 & 0.108 & $<0.001$ & 0.284 & 0.004 & 0.281 & 0.004 \\
\midrule
\multicolumn{8}{l}{\textit{Panel B: log(USDC/USDT) volume}}\\
Global post-pre change & -0.109 & 0.034 & 0.002 & 0.057 & 0.056 & 0.053 & 0.042 \\
Regulated-facing post-pre change & 0.428 & 0.054 & $<0.001$ & 0.259 & 0.099 & 0.255 & 0.093 \\
Regulated-facing $-$ Global DiD & 0.537 & 0.064 & $<0.001$ & 0.265 & 0.043 & 0.261 & 0.039 \\
\bottomrule
\end{tabular}

\par\vspace{0.35em}
\begin{minipage}{0.98\linewidth}
\footnotesize \textit{Notes.} Panel A outcome is the 14-day smoothed USDC share; Panel B outcome is the 14-day smoothed $\log(1+\text{USDC volume})-\log(1+\text{USDT volume})$. Both are standardized by each exchange's pre-April-1 mean and standard deviation and detrended using pre-event observations; estimates are in pre-event standard-deviation units. Regression as in Equation~(\ref{eq:did}).
\end{minipage}
\end{threeparttable}
\end{table}

Exact randomization inference confirms both results (Appendix Table~\ref{tab:app_permutation}): the two-sided $p$-values are 0.006 for the USDC share and 0.021 for relative volume.

Finally, we ask whether the reallocation has an effect on the relative dollar price of USDC and USDT. Using daily low prices against the dollar, the outcome is USDT downside peg pressure minus USDC's, so a positive value means USDT trades further below par than USDC. Table~\ref{tab:downside_peg_pressure} reports the difference-in-differences two ways. Panel A is the daily event-study panel of Equation~(\ref{eq:did}), estimated on all exchange-days in the window, with standard errors reported without clustering and with two-way clustering by exchange and date. Panel B instead collapses each exchange to a single post-minus-pre change and compares the average change across the four Regulated-facing and ten Global exchanges, reporting a heteroskedasticity-robust (HC1) standard error alongside exact randomization $p$-values. With only fourteen exchanges and four treated units, asymptotic clustered and robust standard errors are unreliable, so we base inference on the exact randomization $p$-values. The estimates are positive but small---about 0.6 to 2 basis points---and imprecise, because deviations from parity are tiny and the sample is small.

\begin{table}[H]
\centering
\caption{Downside Peg Pressure Around MiCA}
\label{tab:downside_peg_pressure}
\scriptsize
\begin{minipage}{0.98\linewidth}
\textit{Panel A. Daily event-study panel}\\[0.25em]
\begin{tabular*}{\linewidth}{@{\extracolsep{\fill}}lrrrrrr}
\toprule
Specification & DiD & SE non-cl. & $p$ non-cl. & SE 2-way & $p$ 2-way & Obs. \\
\midrule
14-day avg., no deadband (baseline) & 0.599 & 0.433 & 0.168 & 0.506 & 0.238 & 420 \\
14-day avg., 5 bps deadband & 0.432 & 0.381 & 0.258 & 0.367 & 0.240 & 420 \\
Daily, no deadband & 1.833 & 2.240 & 0.414 & 1.755 & 0.297 & 420 \\
Daily, 5 bps deadband & 1.983 & 2.088 & 0.343 & 1.567 & 0.207 & 420 \\
Daily, 10 bps deadband & 2.092 & 1.884 & 0.268 & 1.426 & 0.143 & 420 \\
\bottomrule
\end{tabular*}

\vspace{0.75em}
\textit{Panel B. Exchange-level post-minus-pre collapse}\\[0.25em]
\begin{tabular*}{\linewidth}{@{\extracolsep{\fill}}lrrrrrrr}
\toprule
Specification & Global $\Delta$ & Reg.-facing $\Delta$ & DiD & HC1 SE & $p$ HC1 & Exact $p$ (2s) & Exact $p$ (+) \\
\midrule
14-day avg., no deadband (baseline) & 0.444 & 1.042 & 0.599 & 0.587 & 0.328 & 0.508 & 0.259 \\
14-day avg., 5 bps deadband & 0.215 & 0.647 & 0.432 & 0.431 & 0.336 & 0.414 & 0.198 \\
Daily, no deadband & 0.978 & 2.811 & 1.833 & 1.590 & 0.271 & 0.499 & 0.233 \\
Daily, 5 bps deadband & 0.216 & 2.199 & 1.983 & 1.249 & 0.138 & 0.272 & 0.076 \\
Daily, 10 bps deadband & -0.457 & 1.634 & 2.092 & 1.168 & 0.099 & 0.193 & 0.038 \\
\bottomrule
\end{tabular*}
\vspace{0.35em}
\footnotesize \textit{Notes.} The outcome is USDT downside peg pressure minus USDC downside peg pressure, in basis points. Downside pressure is $\max\{0,-\log(P^{low})-\tau\}$, where $P^{low}$ is the daily low price against the dollar and $\tau$ is the deadband shown in the row label (0, 5, or 10 basis points). The deadband is an ignored zone around parity: discount deviations smaller than $\tau$ are set to zero before measuring pressure. Positive coefficients mean that USDT becomes more discounted relative to USDC on Regulated-facing exchanges after April 1, 2025. Panel A estimates the daily event-study panel over the 15 days before and after April 1 with exchange fixed effects; standard errors are reported without clustering and with two-way clustering by exchange and date. Panel B first collapses each exchange to one post-minus-pre change and then compares the average change for Regulated-facing and Global exchanges; it reports a heteroskedasticity-robust (HC1) standard error and its $p$-value together with exact randomization $p$-values. The exact $p$-values enumerate all $\binom{14}{4}=1001$ assignments of the four Regulated-facing labels among the 14 exchanges and locate the estimate in that exact permutation distribution: Exact $p$ (2s) is two-sided, and Exact $p$ (+) is the one-sided probability of a difference at least as positive as the one observed. With only 14 exchanges and four treated units, the asymptotic clustered (Panel A) and HC1 (Panel B) standard errors are unreliable, so the exact randomization $p$-values are the preferred basis for inference.
\end{minipage}
\end{table}

\clearpage
\begin{spacing}{1.5}
\bibliographystyle{micaabbrvnat}
\bibliography{refs}
\end{spacing}

\clearpage
\appendix
\setcounter{figure}{0}
\setcounter{table}{0}
\renewcommand{\thefigure}{A\arabic{figure}}
\renewcommand{\thetable}{A\arabic{table}}

% ============================================================
% REVISED ONLINE APPENDIX (consolidated)
% Drop-in replacement for everything after \appendix in the main file.
% Preserved labels referenced from the main text:
%   app:mica_details, tab:app_exchange_chars, tab:app_delistings,
%   app:pretrends, tab:app_volume_legs, tab:app_permutation
% Merged tables are hand-written from current estimates; regenerate
% from the analysis pipeline before submission to keep numbers in sync.
% ============================================================

\section{Online Appendix}
\label{sec:online_appendix}

\subsection{MiCA Institutional Detail}
\label{app:mica_details}

MiCA regulates two broad groups: issuers of crypto-assets and crypto-asset service providers (CASPs). Issuers face disclosure obligations through crypto-asset white papers and rules on fair, clear, and non-misleading communications. CASPs (trading, custody, execution, and transfer services) must obtain authorization and comply with governance, prudential, and conduct requirements; once authorized, they can passport services across the EU (\citealp{EuropeanParliamentCouncil2023MiCA}; Patti, 2024). MiCA's perimeter is not universal: NFTs (in many cases), fully decentralized DeFi activity, and CBDCs sit outside or at the margins \citep{ESMA2025MiCA}.

Legal work stresses the delineation between crypto-assets under MiCA and instruments under existing EU securities law, and the similarity between MiCA's white-paper regime and prospectus-style disclosure (Zetzsche et al., 2021). Recent ECB research details the ART/EMT split, the transitional regime for CASPs, and the reserve architecture for EMTs (Altavilla et al., 2026); the IMF stresses that stablecoin stability depends on reserve quality and enforceable redemption rights (International Monetary Fund, 2025).

\subsection{Exchange Sample}
\label{app:exchange_descriptives}

Table~\ref{tab:app_exchange_chars} reports the exchange-level characteristics used to construct the sample, classify venues, and define the exchange-specific event dates used in robustness checks. The March 31, 2025 delisting wave motivates the April 1, 2025 common event date used in the main analysis; when an exchange-specific date is missing, the alternative-event-date specifications use the common cutoff. %Table~\ref{tab:app_descriptive_usdc_share} provides descriptive pre- and post-MiCA USDC-share comparisons; because the share is computed over combined USDT--USDC volume, the corresponding USDT shares are the complements and are omitted.

\begin{table}[H]
\centering
\begin{threeparttable}
\caption{Exchange Sample, MiCA Compliance, and Event Dates}
\label{tab:app_exchange_chars}
\label{tab:app_delistings}
\scriptsize
\setlength{\tabcolsep}{3pt}
\begin{tabular*}{\linewidth}{@{\extracolsep{\fill}}l r r r c l l l}
\toprule
Exchange
  & \multicolumn{1}{c}{CMC}
  & \multicolumn{1}{c}{Volume}
  & \multicolumn{1}{c}{Visits}
  & MiCA
  & License
  & Jurisdiction
  & USDT Delist. \\
  & \multicolumn{1}{c}{Rank}
  & \multicolumn{1}{c}{(\$bn/day)}
  & \multicolumn{1}{c}{(mn/wk)}
  &
  & Date
  &
  & (EEA) \\
\midrule
\multicolumn{8}{l}{\textit{Panel A: US Exchanges}} \\[2pt]
Coinbase     &  2 &  2.72 &  0.04 & Yes & 2025-06-20 & Luxembourg & 2024-12-13 \\
Gemini       & 24 &  0.22 &  0.37 & Yes & 2025-04-21 & Malta      & 2025-02-28 \\
\addlinespace
\multicolumn{8}{l}{\textit{Panel B: US/EU Exchanges}} \\[2pt]
Kraken       & 14 &  1.27 &  1.61 & Yes & 2025-06-26 & Ireland    & 2025-03-31 \\
Bitstamp     & 18 &  0.52 &  0.14 & Yes & 2025-05-16 & Luxembourg & 2025-01-31 \\
\addlinespace
\multicolumn{8}{l}{\textit{Panel C: Global Exchanges}} \\[2pt]
Binance      &  1 & 20.31 & 11.30 & No  & ---        & ---        & 2025-03-31 \\
Bybit        &  3 &  3.89 &  3.33 & Yes & 2025-05-29 & Austria    & ---        \\
OKX          &  5 &  3.48 &  4.61 & Yes & 2025-01-27 & Malta      & 2024-03-18 \\
Bitget       &  6 &  1.83 &  3.94 & No  & ---        & ---        & ---        \\
Gate.io      &  7 &  3.70 &  4.42 & No  & 2025-09-30 & ---        & 2025-05-11 \\
KuCoin       &  8 &  5.99 &  4.35 & Yes & 2025-11-28 & Austria    & 2025-05-11 \\
MEXC         &  9 &  3.64 &  5.08 & No  & ---        & ---        & ---        \\
HTX          & 10 &  2.96 &  4.67 & No  & ---        & ---        & 2025-05-11 \\
Bitfinex     & 12 &  0.41 &  0.08 & No  & ---        & ---        & ---        \\
BitMart      & 16 &  2.68 &  3.76 & No  & ---        & ---        & ---        \\
\bottomrule
\end{tabular*}
\par\vspace{0.35em}
\begin{minipage}{0.98\linewidth}
\footnotesize \textit{Notes.} The table describes the 14 exchanges in the sample, grouped by audience segment. CMC Rank is the CoinMarketCap ranking as of data collection. Volume is mean 24-hour trading volume in USD billions over the sample period. Visits is the weekly unique-visitor count in millions from CoinMarketCap. MiCA is the compliance classification in the exchange metadata; License Date is the first CASP authorization date recorded in ESMA's official register when available. USDT Delist. (EEA) is the effective date on which the exchange removed USDT pairs for EEA users. Missing event dates, denoted by ---, are replaced by the common April 1, 2025 cutoff in the alternative-event-date specifications. Audience classification is based on Similarweb web-traffic data. Regulated-facing exchanges (Panels A and B) have an EU audience share above 10 percent, and each also has a US audience share above 10 percent; within this group, Panel A (US) has a US audience share above 50 percent and Panel B (US/EU) has a US share between 10 and 50 percent. Global exchanges (Panel C) have both EU and US audience shares below 10 percent. HTX was formerly known as Huobi Pro.
\end{minipage}
\end{threeparttable}
\end{table}

% \begin{table}[H]
% \centering
% \begin{threeparttable}
% \caption{USDC Market Share by Exchange: Pre- and Post-MiCA}
% \label{tab:app_descriptive_usdc_share}
% \scriptsize
% \input{figures/table1_usdc_share.tex}
% \par\vspace{0.35em}
% \begin{minipage}{0.98\linewidth}
% \footnotesize \textit{Notes.} USDC Share is USDC trading volume divided by USDC plus USDT trading volume; the corresponding USDT share is the complement. Volume is average daily trading volume in billions of USD. MiCA indicates regulatory compliance status in the descriptive table classification. Audience is based on Similarweb traffic composition. Pre-MiCA is the 90 days before delisting; Post-MiCA is the 90 days starting at the delisting date. Coinbase USD is treated as USDC. Significance levels: *** $p<0.01$, ** $p<0.05$, * $p<0.10$ from a two-tailed Welch t-test on pre/post percentage-point differences.
% \end{minipage}
% \end{threeparttable}
% \end{table}

\subsection{Aggregate Series and Cross-Sectional Figures}

\begin{figure}[H]
    \centering
    \includegraphics[width=0.9\linewidth]{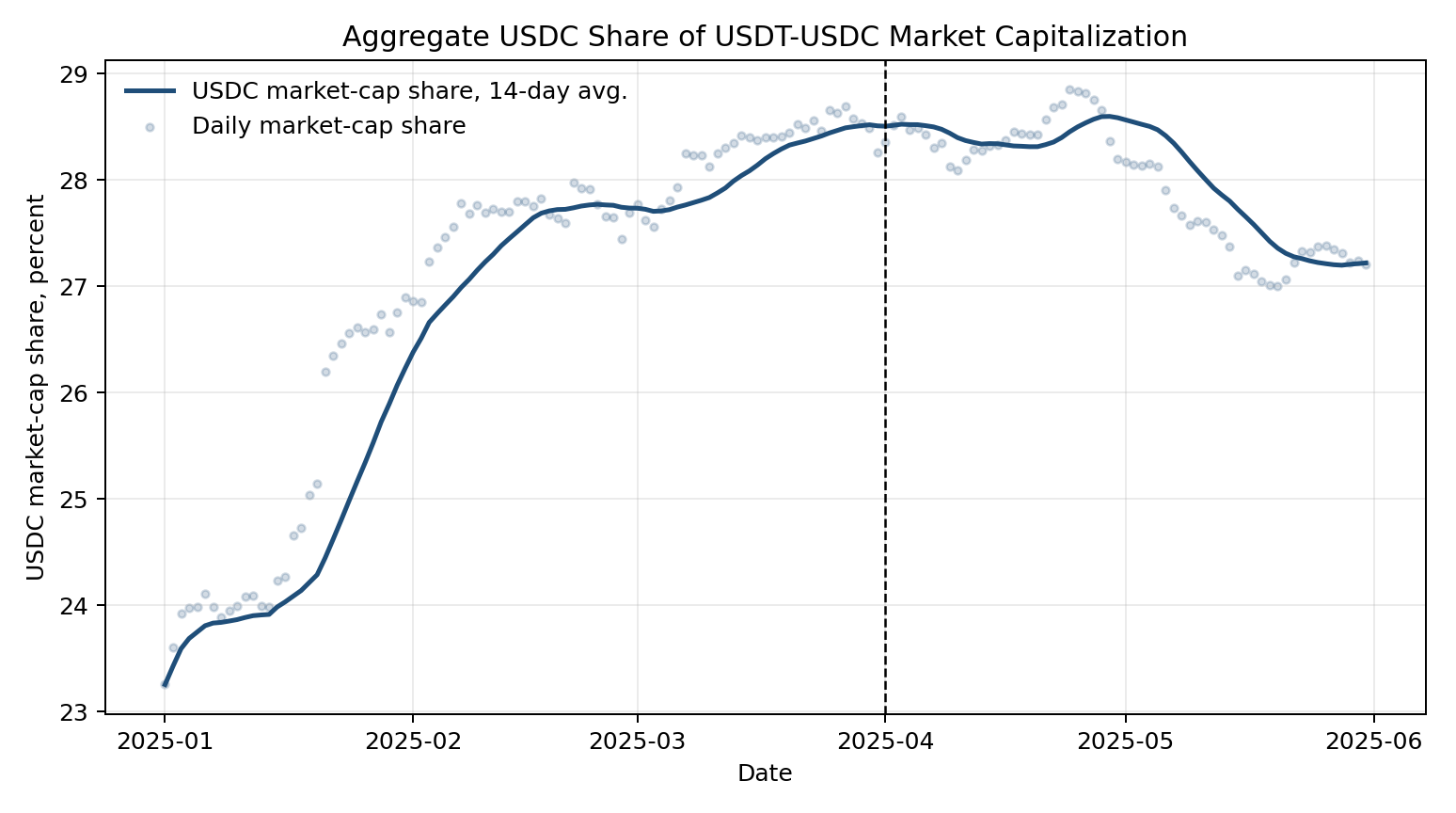}
    \caption{Aggregate USDC Market-Capitalization Share Around April 1, 2025}
    \label{fig:app_market_cap_share}
    \begin{center}
    \parbox{0.95\linewidth}{\scriptsize \textit{Note:} Daily USDC market-capitalization share is USDC market capitalization divided by USDC plus USDT market capitalization, using Coin Metrics data. This figure plots a relative market-capitalization share, not relative trading volume. The line is a 14-day trailing average; dots are daily observations. The vertical dashed line marks April 1, 2025. Values are in percent.}
    \end{center}
\end{figure}

\begin{figure}[H]
    \centering
    \includegraphics[width=0.9\linewidth]{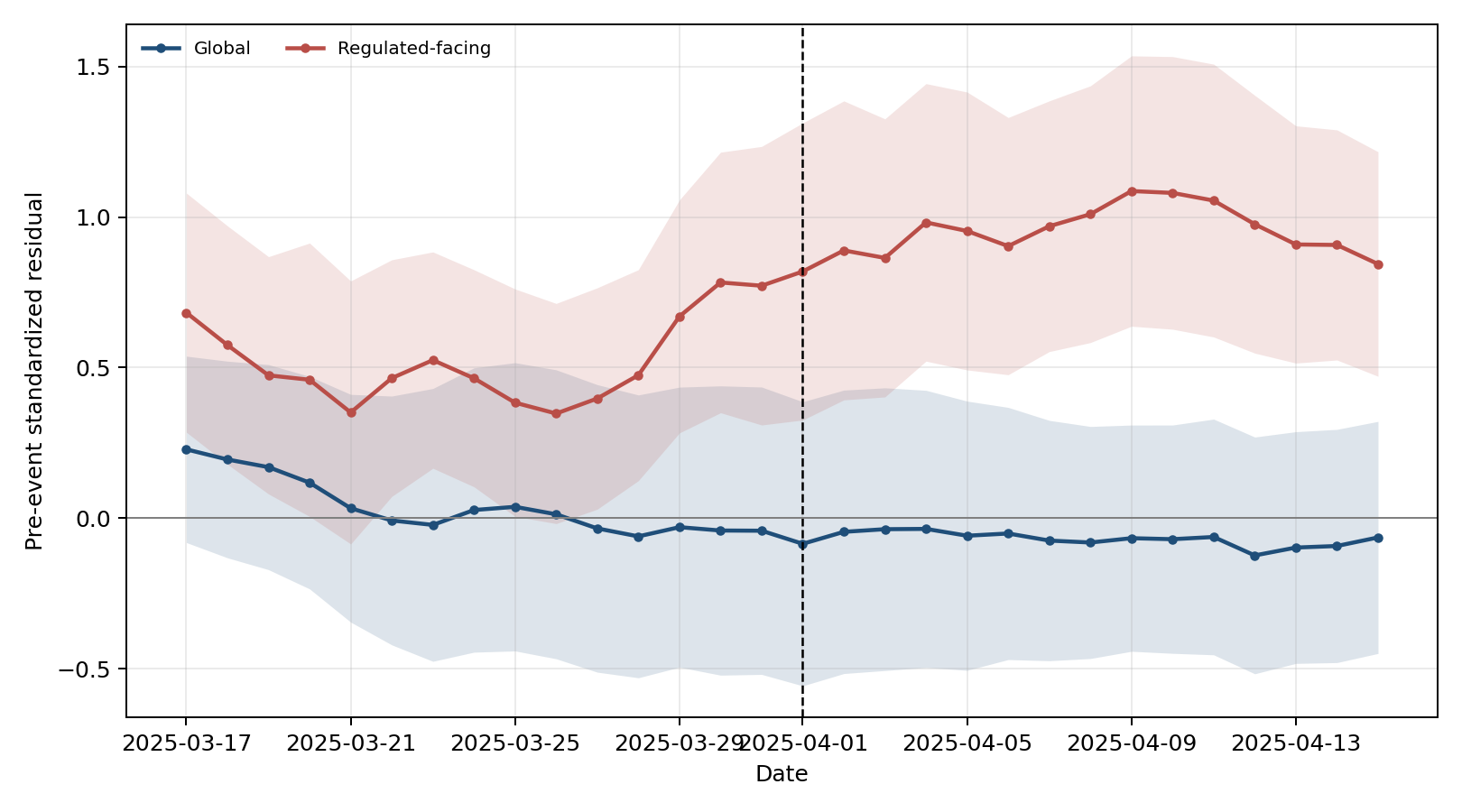}
    \caption{Relative USDC-minus-USDT Volume Around April 1, 2025}
    \label{fig:app_volume_path}
    \begin{center}
    \parbox{0.95\linewidth}{\scriptsize \textit{Note:} Daily equal-weighted Global and Regulated-facing means of the 14-day smoothed $\log(1+\text{USDC volume})-\log(1+\text{USDT volume})$, after exchange-level pre-event standardization and detrending. The vertical dashed line marks April 1, 2025. Shaded bands are $\pm 1$ cross-exchange standard error of the daily group mean.}
    \end{center}
\end{figure}

Figure~\ref{fig:app_market_cap_share} plots USDC's aggregate share of combined USDT--USDC market capitalization using Coin Metrics data. This is a relative market-capitalization measure, not a trading-volume measure; it is reported only as an aggregate descriptive check. Figure~\ref{fig:app_volume_path} shows the cross-sectional path of the relative-volume outcome, the volume analogue of main-text Figure~\ref{fig:share_path_event}: the series is flat to slightly declining on Global exchanges and rises on Regulated-facing exchanges.

\subsection{Raw Pre-Event Trends}
\label{app:pretrends}

\begin{figure}[H]
    \centering
    \includegraphics[width=0.9\linewidth]{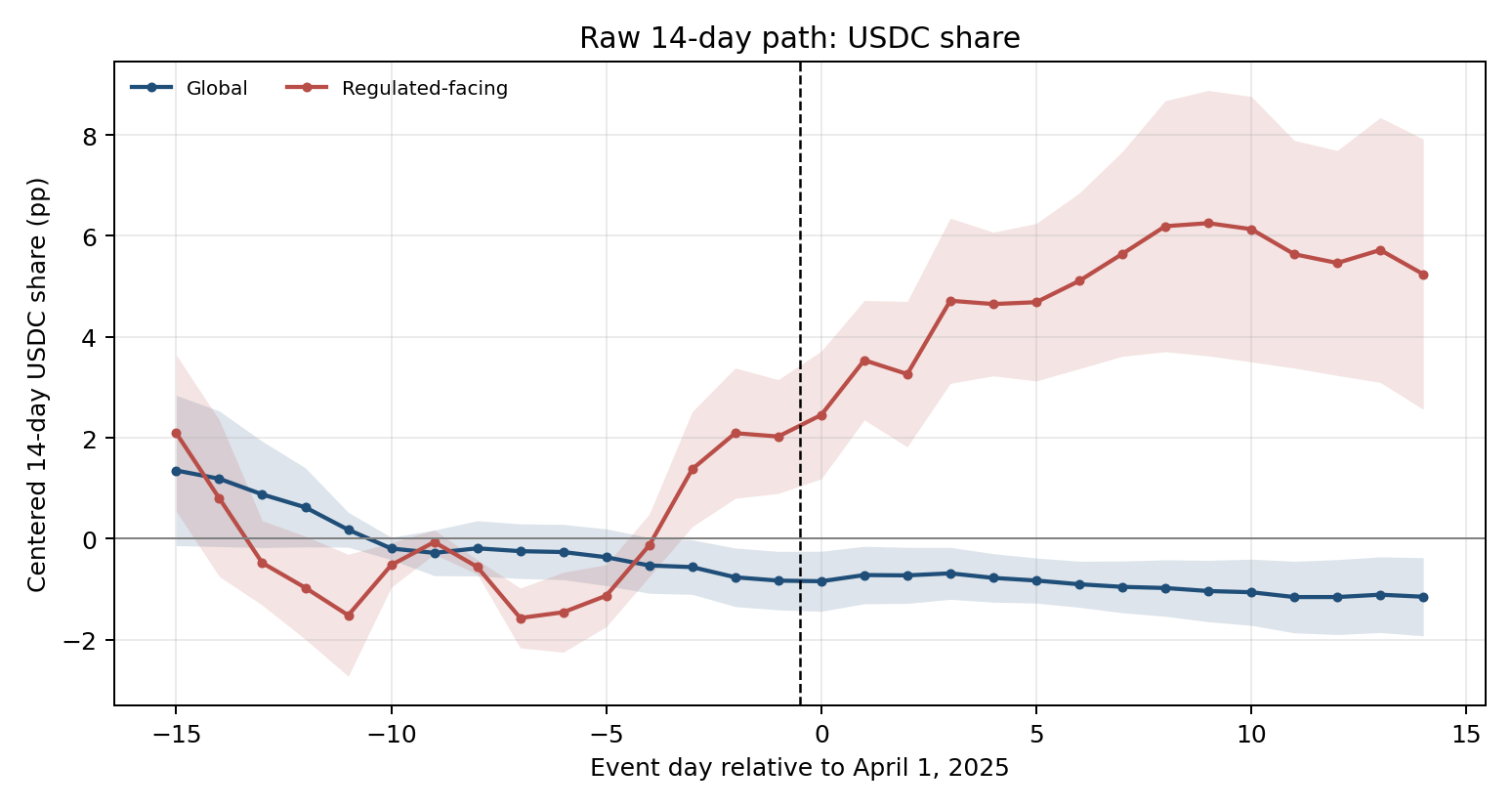}
    \caption{Raw Pre-Event Path: USDC Share}
    \label{fig:app_pretrend_share}
    \begin{center}
    \parbox{0.95\linewidth}{\scriptsize \textit{Note:} 14-day trailing-average USDC share in percentage points, without standardization or detrending, centered on each exchange's pre-April-1 mean and averaged within group. Shaded bands are $\pm 1$ cross-exchange standard error of the daily group mean. The vertical dashed line marks April 1, 2025.}
    \end{center}
\end{figure}

\begin{figure}[H]
    \centering
    \includegraphics[width=0.9\linewidth]{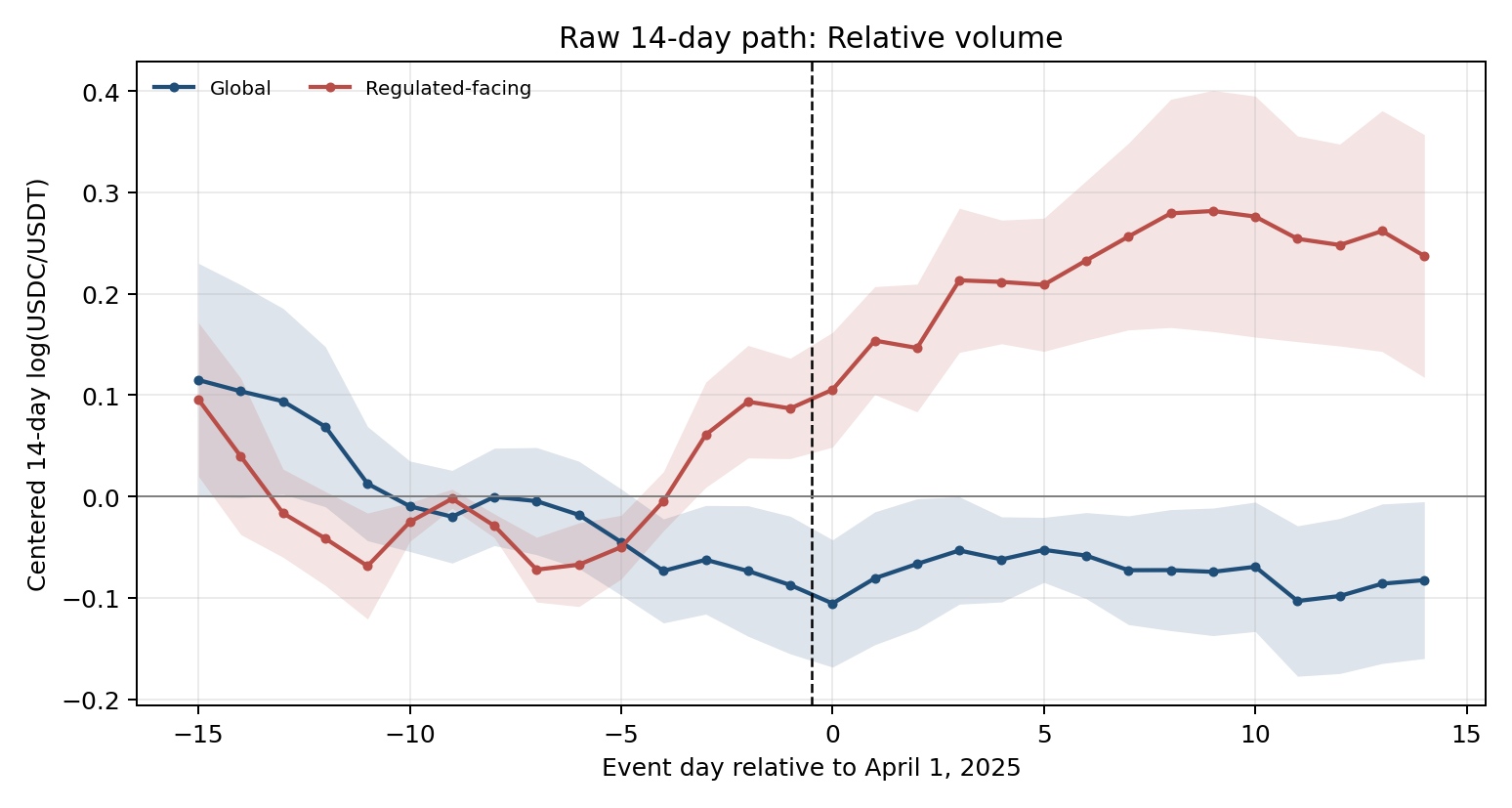}
    \caption{Raw Pre-Event Path: Relative Volume}
    \label{fig:app_pretrend_volume}
    \begin{center}
    \parbox{0.95\linewidth}{\scriptsize \textit{Note:} 14-day trailing-average $\log(1+\text{USDC volume})-\log(1+\text{USDT volume})$, without standardization or detrending, centered on each exchange's pre-April-1 mean and averaged within group. Shaded bands are $\pm 1$ cross-exchange standard error of the daily group mean. The vertical dashed line marks April 1, 2025.}
    \end{center}
\end{figure}

Figures~\ref{fig:app_pretrend_share} and~\ref{fig:app_pretrend_volume} plot the two outcomes \emph{without} the exchange-level standardization or detrending used in the main specification. Each exchange's 14-day trailing average is centered on its pre-event mean, and we plot equal-weighted Global and Regulated-facing group means over the event window. In both cases the two groups move together before April 1 and diverge only around the cutoff, with Regulated-facing rising. A binned dynamic difference-in-differences on the raw daily series does not reject parallel pre-trends (joint $p$-values of 0.12 for the share and 0.15 for relative volume), although the daily series are noisy. The standardization and detrending in the main specification therefore sharpen precision rather than create the effect.

\subsection{Volume Decomposition: USDC and USDT Legs}
\label{app:volume_legs}

The relative-volume result in Panel B of Table~\ref{tab:main_event} moves toward USDC, but the ratio alone cannot say whether USDC trading expands or USDT trading contracts. Table~\ref{tab:app_volume_legs} decomposes it by running the same event study separately on each leg. The shift is driven by USDT: on Regulated-facing exchanges relative to Global exchanges, USDT trading volume falls by about 20 percent (exchange-clustered $p{=}0.003$), while USDC volume does not change significantly. This is consistent with the delisting mechanism---where USDT is removed for European users, its trading contracts---rather than with an expansion of USDC.

\begin{table}[H]
\centering
\begin{threeparttable}
\caption{Volume Decomposition: USDC and USDT Legs}
\label{tab:app_volume_legs}
\small
\setlength{\tabcolsep}{4pt}
\begin{tabular}{lrrrrrr}
\toprule
Outcome & DiD (log pts) & DiD (\%) & SE exch. & $p$ exch. & SE 2-way & $p$ 2-way \\
\midrule
USDC trading volume & $+3.9$ & $+3.9$ & 11.092 & 0.728 & 10.773 & 0.721 \\
USDT trading volume & $-22.5$ & $-20.1$ & 7.675 & 0.003 & 7.731 & 0.004 \\
\bottomrule
\end{tabular}

\par\vspace{0.35em}
\begin{minipage}{0.98\linewidth}
\footnotesize \textit{Notes.} Each row is a difference-in-differences (Regulated-facing minus Global) from the baseline event study of Equation~(\ref{eq:did}) run on a single leg of volume. The outcome is the 14-day smoothed $\log(1+\text{volume})$ of USDC or of USDT, detrended by an exchange-specific pre-event linear time trend. The specification is unstandardized, so the two legs sum to the relative-volume outcome of Panel~B in Table~\ref{tab:main_event}: the USDC leg minus the USDT leg equals the relative-volume difference-in-differences. DiD is reported in log points and in exact percent; standard errors are exchange-clustered and two-way (exchange and date) clustered.
\end{minipage}
\end{threeparttable}
\end{table}

\subsection{Robustness: Smoothing Windows and Leave-One-Out}
\label{app:robustness_core}

Table~\ref{tab:app_smoothing_combined} reports the difference-in-differences for both outcomes under alternative trailing smoothing windows (Panels A and B) and for the raw percentage-point USDC-share specification (Panel C). For each window, standardization uses the exchange's own pre-event mean and standard deviation of that smoothed series. The 14-day baseline reproduces the main-text estimates, and the raw percentage-point difference-in-differences is 5.92 points (exchange-clustered $p{<}0.001$).

\begin{table}[htbp]
\centering
\begin{threeparttable}
\caption{Smoothing-Window and Raw-Share Robustness}
\label{tab:app_smoothing_combined}
\small
\setlength{\tabcolsep}{4pt}
\begin{tabular}{lrrrcccc}
\toprule
Specification & Global & Reg.-facing & DiD & $p$ non-cl. & $p$ exch. & $p$ 2-way & Signs \\
\midrule
\multicolumn{8}{l}{\textit{Panel A: USDC share (pre-event standard-deviation units)}} \\[2pt]
1-day               & $-0.141$ & 0.129 & 0.270 & 0.176      & 0.291 & 0.352 & Yes \\
7-day               & $-0.335$ & 0.569 & 0.904 & $<0.001$   & 0.010 & 0.008 & Yes \\
14-day (baseline)   & $-0.243$ & 0.576 & 0.818 & $<0.001$   & 0.004 & 0.004 & Yes \\
30-day              & $-0.206$ & 0.109 & 0.315 & $<0.001$   & 0.199 & 0.191 & Yes \\
\addlinespace
\multicolumn{8}{l}{\textit{Panel B: $\log(\text{USDC}/\text{USDT})$ volume (pre-event standard-deviation units)}} \\[2pt]
1-day               & $-0.006$ & 0.046 & 0.052 & 0.697      & 0.754 & 0.775 & Yes \\
7-day               & $-0.062$ & 0.398 & 0.460 & $<0.001$   & 0.060 & 0.059 & Yes \\
14-day (baseline)   & $-0.109$ & 0.428 & 0.537 & $<0.001$   & 0.043 & 0.039 & Yes \\
30-day              & $-0.263$ & 0.056 & 0.319 & $<0.001$   & 0.198 & 0.194 & Yes \\
\addlinespace
\multicolumn{8}{l}{\textit{Panel C: raw USDC share, percentage points}} \\[2pt]
14-day smoothed     & $-0.94$  & 4.98  & 5.92  & $<0.001$   & $<0.001$ & $<0.001$ & Yes \\
No smoothing        & $-0.34$  & 3.21  & 3.55  & 0.014      & 0.128 & 0.198 & Yes \\
\bottomrule
\end{tabular}
\par\vspace{0.35em}
\begin{minipage}{0.98\linewidth}
\footnotesize \textit{Notes.} Panels A and B standardize each exchange by its pre-April-1 mean and standard deviation of the smoothed series and detrend on pre-event observations; Panel C is the raw USDC share in percentage points. The Signs column indicates whether Global decreases, Regulated-facing increases, and the Regulated-facing-minus-Global DiD is positive.
\end{minipage}
\end{threeparttable}
\end{table}

Table~\ref{tab:app_loo_combined} repeats the baseline 14-day specification for both outcomes after omitting one exchange at a time. The USDC-share DiD stays positive and significant at the 5 percent level in all fourteen omissions (exchange-clustered $p$ between $<0.001$ and 0.037); the relative-volume DiD remains positive in all omissions.

\begin{table}[htbp]
\centering
\begin{threeparttable}
\caption{Leave-One-Out Robustness: USDC Share and Relative Volume}
\label{tab:app_loo_combined}
\scriptsize
\setlength{\tabcolsep}{5pt}
\begin{tabular}{ll rcc rcc}
\toprule
 & & \multicolumn{3}{c}{USDC share} & \multicolumn{3}{c}{$\log(\text{USDC}/\text{USDT})$ volume} \\
\cmidrule(lr){3-5} \cmidrule(lr){6-8}
Omitted & Group & DiD & $p$ exch. & $p$ 2-way & DiD & $p$ exch. & $p$ 2-way \\
\midrule
Binance  & Global           & 0.850 & 0.003    & 0.003    & 0.545 & 0.042 & 0.038 \\
BitMart  & Global           & 0.846 & 0.003    & 0.003    & 0.564 & 0.034 & 0.031 \\
Bitfinex & Global           & 0.750 & 0.007    & 0.007    & 0.523 & 0.050 & 0.048 \\
Bitget   & Global           & 0.784 & 0.006    & 0.005    & 0.502 & 0.058 & 0.054 \\
Bitstamp & Regulated-facing & 0.723 & 0.037    & 0.033    & 0.458 & 0.175 & 0.168 \\
Bybit    & Global           & 0.860 & 0.003    & 0.002    & 0.570 & 0.032 & 0.029 \\
Coinbase & Regulated-facing & 0.619 & 0.024    & 0.023    & 0.307 & 0.146 & 0.133 \\
Gate     & Global           & 0.814 & 0.005    & 0.005    & 0.531 & 0.047 & 0.043 \\
Gemini   & Regulated-facing & 0.875 & 0.015    & 0.014    & 0.670 & 0.031 & 0.028 \\
HTX      & Global           & 0.816 & 0.005    & 0.005    & 0.524 & 0.050 & 0.046 \\
Kraken   & Regulated-facing & 1.056 & $<0.001$ & $<0.001$ & 0.711 & 0.011 & 0.010 \\
KuCoin   & Global           & 0.845 & 0.003    & 0.003    & 0.541 & 0.043 & 0.039 \\
MEXC     & Global           & 0.769 & 0.007    & 0.006    & 0.530 & 0.048 & 0.044 \\
OKX      & Global           & 0.848 & 0.003    & 0.003    & 0.542 & 0.043 & 0.039 \\
\bottomrule
\end{tabular}
\par\vspace{0.35em}
\begin{minipage}{0.98\linewidth}
\footnotesize \textit{Notes.} Each row omits one exchange and reruns the baseline 14-day pre-event-standardized event study for each outcome. DiD is the Regulated-facing-minus-Global difference-in-differences in pre-event standard-deviation units. Both DiDs remain positive across all omissions.
\end{minipage}
\end{threeparttable}
\end{table}

\subsection{Small-Sample Inference: Wild Bootstrap and Exact Randomization}
\label{app:inference}

With 14 exchanges and four treated, cluster-robust $p$-values rest on few clusters. Table~\ref{tab:app_permutation} therefore reports, for each outcome, the Rademacher wild cluster bootstrap $p$-value (999 draws, clustering by exchange) and exact randomization inference: we compute each exchange's post-minus-pre change in the standardized outcome, enumerate all $\binom{14}{4}=1001$ assignments of the treated group, and report the share of assignments whose Regulated-facing-minus-Global difference is at least as large in absolute value as the observed one.

\begin{table}[htbp]
\centering
\begin{threeparttable}
\caption{Small-Sample Inference: Wild Cluster Bootstrap and Exact Randomization}
\label{tab:app_permutation}
\small
\begin{tabular}{lrcccc}
\toprule
Outcome & DiD & Wild $p$ & Exact $p$ (2s) & Exact $p$ (1s) & Assignments \\
\midrule
USDC share                          & 0.818 & 0.053 & 0.006 & 0.006 & 1001 \\
$\log(\text{USDC}/\text{USDT})$ volume & 0.537 & 0.110 & 0.021 & 0.021 & 1001 \\
\bottomrule
\end{tabular}
\par\vspace{0.35em}
\begin{minipage}{0.98\linewidth}
\footnotesize \textit{Notes.} DiD is the observed Regulated-facing-minus-Global difference-in-differences for the baseline 14-day pre-event-standardized specification, in pre-event standard-deviation units. Wild $p$ is from a Rademacher wild cluster bootstrap with exchange-level weights and 999 replications. Exact $p$-values are from randomization inference over all $\binom{14}{4}=1001$ assignments of the four Regulated-facing labels, based on exchange-level post-minus-pre changes; (2s) is two-sided and (1s) one-sided.
\end{minipage}
\end{threeparttable}
\end{table}

\subsection{Alternative Event Dates}
\label{app:alt_event_dates}

The main analysis uses a common April 1, 2025 cutoff. Table~\ref{tab:app_alt_event_dates} instead assigns each exchange its own event date---the USDT delisting date in Panels A and B, and the MiCA license date in Panel C---when available, and April 1, 2025 otherwise (dates in Table~\ref{tab:app_delistings}); standardization and detrending use observations before that exchange's cutoff. Both outcomes keep the positive DiD sign under delisting-date cutoffs, but clustered inference is less precise, as expected, because spreading the event across exchange-specific dates dilutes the sharp common-date variation. License dates are less directly tied to the asset-menu mechanism than delisting dates, so Panel C is mainly a check on event timing.

\begin{table}[htbp]
\centering
\begin{threeparttable}
\caption{Alternative Event-Date Cutoffs}
\label{tab:app_alt_event_dates}
\small
\setlength{\tabcolsep}{4pt}
\begin{tabular}{lrrrrrrr}
\toprule
Effect & Estimate & SE non-cl. & $p$ non-cl. & SE exch. & $p$ exch. & SE 2-way & $p$ 2-way \\
\midrule
\multicolumn{8}{l}{\textit{Panel A: USDC share, delisting-date cutoffs}} \\[2pt]
Global post-pre change            & $-0.316$ & 0.066 & $<0.001$ & 0.107 & 0.003 & 0.103 & 0.002 \\
Regulated-facing post-pre change  & 0.128    & 0.104 & 0.218    & 0.285 & 0.652 & 0.285 & 0.653 \\
Regulated-facing $-$ Global DiD   & 0.445    & 0.123 & $<0.001$ & 0.304 & 0.144 & 0.303 & 0.142 \\
\addlinespace
\multicolumn{8}{l}{\textit{Panel B: $\log(\text{USDC}/\text{USDT})$ volume, delisting-date cutoffs}} \\[2pt]
Global post-pre change            & $-0.172$ & 0.041 & $<0.001$ & 0.077 & 0.026 & 0.075 & 0.021 \\
Regulated-facing post-pre change  & 0.044    & 0.064 & 0.499    & 0.203 & 0.830 & 0.203 & 0.830 \\
Regulated-facing $-$ Global DiD   & 0.216    & 0.076 & 0.005    & 0.217 & 0.321 & 0.215 & 0.317 \\
\addlinespace
\multicolumn{8}{l}{\textit{Panel C: $\log(\text{USDC}/\text{USDT})$ volume, license-date cutoffs}} \\[2pt]
Global post-pre change            & $-0.022$ & 0.038 & 0.557    & 0.125 & 0.859 & 0.124 & 0.858 \\
Regulated-facing post-pre change  & 0.054    & 0.059 & 0.358    & 0.099 & 0.583 & 0.097 & 0.577 \\
Regulated-facing $-$ Global DiD   & 0.077    & 0.070 & 0.276    & 0.160 & 0.631 & 0.157 & 0.626 \\
\bottomrule
\end{tabular}
\par\vspace{0.35em}
\begin{minipage}{0.98\linewidth}
\footnotesize \textit{Notes.} Each exchange uses its own USDT delisting date (Panels A and B) or MiCA license date (Panel C) when available and April 1, 2025 otherwise; the dates are listed in Table~\ref{tab:app_delistings}. Standardization and detrending use observations before that exchange's cutoff. Estimates are in pre-event standard-deviation units.
\end{minipage}
\end{threeparttable}
\end{table}

\clearpage
\subsection*{Appendix References}
\begin{list}{}{%
  \setlength{\leftmargin}{1.5em}
  \setlength{\itemindent}{-1.5em}
  \setlength{\parsep}{2pt}
  \setlength{\itemsep}{4pt}}
\item Altavilla, C., M. Boucinha, L. Burlon, R. Adalid, R. Fortes, and F. Maruhn (2026). Stablecoins and Monetary Policy Transmission. ECB Working Paper Series 3199, European Central Bank.
\item International Monetary Fund (2025). Understanding Stablecoins. Departmental Paper, International Monetary Fund, Washington, DC.
\item Patti, F. (2024). The European MiCA Regulation: A New Era for Initial Coin Offerings. \textit{Georgetown Journal of International Law} 55.
\item Zetzsche, D. A., F. Annunziata, D. W. Arner, and R. P. Buckley (2021). The Markets in Crypto-Assets Regulation (MiCA) and the EU Digital Finance Strategy. \textit{Capital Markets Law Journal} 16(2), 203--225.
\end{list}

\end{document}